\documentstyle[11pt]{article}
\input epsf

\newcommand{\nl}{\newline}

\newcommand{\eps}{{\epsilon}}

\newcommand{\beq}{\begin{equation}}
\newcommand{\eeq}{\end{equation}}
\newcommand{\beqa}{\begin{eqnarray}}
\newcommand{\eeqa}{\end{eqnarray}}
\newcommand{\ben}{\begin{enumerate}}
\newcommand{\een}{\end{enumerate}}
\newcommand{\bi}{\begin{itemize}}
\newcommand{\ei}{\end{itemize}}

\newcommand{\lsim}{\hbox{ {\raisebox{0.06cm}{$<$} \raisebox{-0.14cm}{$\!\!\!\!\!\!\!\!\: \sim$}} } }
\newcommand{\rsim}{\hbox{ {\raisebox{0.06cm}{$>$} \raisebox{-0.14cm}{$\!\!\!\!\!\!\!\!\: \sim$}} } }
 \title{On the significance of quantum effects and interactions for the
apparent universality of Bloch laws for $M_s(T)$}

\author{U.\ Krey\thanks{e-mail uwe.krey@physik.uni-regensburg.de}
\\Inst.\ of Physics II, Universit\"at Regensburg, D-09040 Regensburg, Germany
} 
\date{v1: March 12, v2=v3: May 27, 2003; accepted by JMMM}
\begin{document}

\large\maketitle 
\begin{abstract}
 
\noindent
 The apparent universality of Bloch's  $T^{3/2}$-law for the
temperature dependence of the spontaneous magnetisation, and of
generalizations thereof, is considered. It is argued that in the derivation
one should not only consider the exchange interaction between the spins, but
also the other interactions between them, leading to elliptical spin
precession and deviations from the parabolic dispersion of magnons. Also
interaction effects are important to explain the {\it apparent} universality of
generalized Bloch law exponents $\eps_B$, defined by
$M_s(T)= M_s(0)-{\rm const.}\cdot T^{\eps_B}$, valid in a wide temperature
range $T_1<T<T_2$, and for dimensionalities $d=1,\,\,2$, and $3$. The
above-mentioned temperature range, the 'Bloch range', lies {\it above} the
quantum range, where magnetic long-range order (e.g.\ in $d=2$ dimensions)
is nontrivially enforced by the additional interactions, but {\it below} the
thermal critical region, where universal 'anomalous scaling dimensions'
apply. In contrast, for the Bloch temperature region, the universality is only apparent,
i.e.\ a crossover-phenomenon, and simple scaling considerations with 'normal
dimensions' apply. However, due to interactions, the Bloch exponent $\eps_B$
depends not only on the dimensionality $d$ of the system, but also on the
spin quantum number $s\,\,\, ({\rm mod} (1/2))$ of the system, i.e.\ for
given $d$ the Bloch exponent $\eps_B$ is different for half-integer $s$ and
for integer $s$. \end{abstract} {\underline{PACS numbers}}: 75. Magnetic
properties;\nl 05.50 Fh Phase Trans\-itions: General Studies ;\nl

{\section{Introduction:}} This paper grew out of discussions following a
recent presentation of certain experimental results in our institute, which
apparently, and unexpectedly, showed the $T^{3/2}$ Bloch law for
$M_s(T)$ for a nanostructured {\it planar} system, \cite{Kipferl}; and
although I originally thought that the considerations presented below
were too simple for publication, some of the part\-i\-cipants of those
discussions suggested that I should write them down. So here I do so,
just hoping to broaden and intensify thereby more personal discussions
at the expense of being necessarily rather informal.

Bloch's famous $T^{3/2}$-law, and the natural generalization thereof, is
written as follows: \beq \label{eq0} M_s(T)=M_s(0)-{\rm const.}\cdot
T^{\eps_B}\,, \eeq where according to the original derivation of F.\
Bloch for a $d=3$-dimensional Heisenberg model, \cite{Bloch}, the Bloch
exponent $\eps_B$ would be $3/2$, while a recent analysis of various
experimental data in different dimensions by U. K\"obler and coworkers,
see Ref.\ \cite{Koebler} and references therein, led to a set of values
to be presented below, which are partially much different from the value
$3/2$, but appear to be quite universal. In particular, the number
$\eps_B=3/2$ does not only follow from the original theoretical analysis
of F.\ Bloch and the later refinement of this analysis by F.J.\ Dyson,
\cite{Dyson}, but (unexpectedly) it was also observed by U.\ Gradmann
{\it et al.} in $d=2$-dimensional ultrathin ferromagnetic films, see e.g.\
Ref.\ \cite{Gradmann1}, and now it is apparently also observed in the planar
nanostructures of Ref.\ \cite{Kipferl}. Moreover, the experimental analysis
of U.\ K\"obler {\it et al.}, Ref.\ \cite{Koebler}, led to the
above-mentioned set of apparently universal numbers for the Bloch exponent
$\eps_B$, where $\eps_B$ depends not only on the dimensionality
$d$ of the ferromagnetic or antiferromagnetic system considered, but
also assumes different values for integer rsp.\ half-integer spin
quantum number $s$: {\it namely $\eps_B=5/2$ rsp.\ $=3/2$ rsp.\ $2$ for
$d=1$ rsp.\ $=2$ rsp.\ $3$ and half-integer $s$; whereas $\eps_B=3$
rsp.\ $=2$ rsp.\ $9/2$ for $d=1$ rsp.\ $=2$ rsp.\ $3$ and integer $s$.}
(The $\eps_B$-values for $d=1$ rsp.\ $=2$ are also observed for {\it
anisotropically} disordered or amorphous $d=2$- rsp.\ $=3$-dimensional
systems.)

Of course, as a theorist, one should always be  skeptical against
experimental data which may be biased by theoretical claims. Therefore, in
the present context it is stressed right from the beginning that the
following presentation is not based on these data, but only {\it motivated}
by it.

 In particular, and intentionally, I always write '{\underline{\it
apparently}} universal' instead of simply 'universal', since the Bloch
behaviour of Eqn.\ (\ref{eq0}) is generally neither observed in the region
of extremely low temperatures near $T=0$, where quantum behaviour dominates,
nor is it observed in the genuine thermal critical region, where one would
obtain scaling with so-called anomalous dimensions, i.e.\ the 'critical
exponents'. Instead, the above-mentioned behaviour should be considered as a
crossover-phenomenon, observed in a (wide) intermediate temperature region
$T_1 < T <T_2$, see below. In particular, the scaling dimensions involved
correspond to simple powers of the basic thermal energy $k_BT$ rsp.\ of
$(k_BT)^{1/2}$, where $k_B$ is Boltzmann's constant. But in agreement with
the fact that magnon-magnon interactions (although not yet thermally
critical) already plays a role in the Bloch region (particularly at $d=2$),
the Bloch exponent, according to the considerations presented below, depends
on the spin quantum number $s$ in the above-mentioned way.

The schedule of the present paper is as follows: In the next chapter the
original Bloch derivation for a $d=3$-dimensional Heisenberg system is
considered, and, for later heuristical purposes, a simplified 'poor man's'
variant thereof. Then I consider planar systems ($d=2$). Next, $d=1$ and
$d=3$ follow, all three cases for $s=1/2$. Finally I consider $s=1$, thereby
collecting the whole set of 'apparent Bloch universality classes' of
K\"obler. At the end, in an additional chapter, the conclusions are
presented.

\section{ Bloch's law for $d=3$} According to this law (which is derived in
any textbook on solid-state magnetism and is one of the most prominent
results of theoretical physics) the temperature dependence of
the magnetisation
$M_s(T)$ of a three-dimensional ferromagnet is  given by
\beq\label{eq1}
M_s(T)\equiv M_s(0)-{\rm const.}\times\int\limits_{BZ}\,{\rm d}^3k\frac{1}{\exp \beta
\eps(\vec k) -1}\,,
\eeq
because each excited magnon reduces the magnetic moment of a ferromagnetic
sample by a Bohr magneton. Here $T$ is the Kelvin temperature, $\beta
=(k_BT)^{-1}$ with the Boltzmann constant $k_B$, and the fraction
$\{\exp (\beta\eps (\vec k))-1\}^{-1}$ represents the thermal expectation
value of the number of magnons with excitation energy $\eps(\vec k)$, where
the wave-vector $\vec k$ has its usual meaning. The integration is over the
Brillouin zone BZ of the crystal.  Since for $k\cdot a \ll 1$
(where $a$ is the lattice constant) the excitation energy  has a
parabolic dispersion, $\eps(\vec k) \propto k^2$, the integral in Eq.\
(\ref{eq1}) converges for $d=3$, whereas for $d=2$ it would be 'infrared
divergent' in agreement with the famous Mermin-Wagner theorem,
\cite{Mermin}.

For the excitation energy $\eps(\vec k)$ of a magnon in a Heisenberg
ferromagnet (and also in itinerant ferromagnets when the spin-orbit
interaction, and also the Stoner excitations, are neglected against the
collective magnon-like spin excitations) one simply has the following
dispersion (we assume cubic symmetry and extend the validity of the formula
over the whole integration region): $\eps(\vec k)=D\cdot k^2$, where $D$ is
the so-called spin-wave stiffness. Therefore, by the substitution $x:=\beta
D \cdot k^2$ and the replacement ${\rm d}^3k=4\pi k^2 {\rm d}k$ one gets the
famous result $M_s(T)=M_s(0)-{\rm const'.}\cdot (\frac{k_BT}{D})^{3/2}$,
where (up to exponentially-small terms) the constants ${\rm const.}$ and
${\rm const.'}$ are related as follows: ${\rm const.'}={\rm const.} \times
2\pi\int\limits_0^\infty\,\frac{x^{1/2}{\rm d}x}{\exp x -1}$.  The analysis
has been refined in the monumental work of F.J.\ Dyson, see Ref.\
\cite{Dyson}, but this is not important below, except at the end.

Rather, for Anderson's ''{\it poor man}'', \cite{Anderson}, instead of the
 usual derivation, one can give the following simpler argument, which
 later-on will also serve for heuristical purposes: $\exp{\beta Dk^2} -1$ is
 approximated for long enough wavelengths and/or high enough $T$ by the
 'quasi-classical thermal-energy approximation' $\beta D\cdot k^2$, so that
 one simply gets $M_s(T)=M_s(0)-{\rm const.}(\frac{k_BT}{D})\times PhsR(T)$,
 where the factor $k_BT$ represents the thermal equipartition energy, while
 $PhsR(T)$ means a typical phase-space radius in k-space \footnote{i.e.\
 $PhsR(T)$ is some kind of reciprocal {\it thermal de Broglie wavelength}},
 replacing the integral $\int\limits_0^{'\pi/a'}\,\frac{4\pi k^2}{k^2}{\rm
 d}k\times g_k(T)$, where $g_k(T)$ is a temperature-dependent dimensionless
 weight-function. Here $'\pi /a'$ represents a (very large)
 wavenumber-cutoff corresponding to the upper edge of the Brillouin zone,
 which is replaced by a sphere as in typical renormalization group
 arguments. But it would be wrong, if at this place one would perform
 directly the integration without any idea of the weight-function (after
 having made the above-mentioned 'equipartition approximation' leading to
 the 'quasi-classical thermal-energy prefactor'
 $(\frac{k_BT}{D})$). Instead, one gets the correct $PhsR(T)$ by a
 simple 'Pippard argument', i.e.\ simply by equating the {\it
 dominating} energy- resp.\ temperature-ranges: $\eps(\vec k_{\rm
 dom.})=D\cdot k_{\rm dom.}^2\equiv k_BT$, \ i.e.\ $PhsR(T)\equiv k_{\rm
 dom.}(T)\,\,(\cong\Delta k(T)) \equiv (\frac{k_BT}{D})^{1/2}$.

\section{Planar systems}
What would be different in a planar system magnetised in-plane~?\nl One
would again expect a quasi-classical 'equipartition approximation'
$\exp\beta \eps(\vec k)-1\to \beta \eps(\vec k)$, but otherwise one would
have a lot of differences and, generally, quite subtle behaviour:
\bi\item First, instead of the simple parabolic dispersion $\eps(\vec
k)=D\cdot k^2$, representing excitations with circular spin precession, one
would have a more general 'square root' formula, $\eps(\vec
k)=\sqrt{\eps_a(\vec k)\cdot\eps_b(\vec k)}$, representing an elliptical
precession with the main axes $a$ and $b$, respectively.

 For example, if one is dealing with a film of infinite extension in the x-
and y-directions, with almost neglegible thickness $t_h$, then (if the film
is magnetised in-plane, e.g.\ in the x-direction, see below) spin-wave
deviations in the z-direction are strongly disfavoured due to the
demagnetising field
$H_z^{\rm DM}=-4\pi M_z$, where $H_z^{\rm DM}$ and $M_z$ are the z-components of the
demagnetising field and the magnetisation, respectively. Thus, one would have
$\eps_b(\vec k)\equiv C_b+D\cdot k^2\,\,(\cong C_b$ in the long-wavelength
limit, where $C_b=\mu_B \cdot 4\pi M_s$ is the effective anisotropy energy
corresponding to the demagnetising field ($\mu_B$ is Bohr's elementary
magnetic moment, the 'magneton'.)). In contrast, spin deviations in the
$y$-direction would not be disfavoured, i.e.\ $\eps_a(\vec k)\cong
D\cdot k^2$, as before (but see the following text, which should make things
more precise):

 Actually, the magnetostatic interactions giving rise to the above-mentioned
demagnetising field $-4\pi M_z$, also produce in-plane field components,
which are proportional to the (small) thickness $t_h$ of the ultrathin film.
(Here, 'ultrathin' means that $t_h$ is of the order-of-magnitude of some
lattice constants; thus, even a monolayer should be approximated in this way
by a continuous film of finite thickness $t_h=a$, and one should not simply
set $t_h=0$, instead.)

\item Second, the in-plane magnetostatic fields are not constant, but in
contrast to the exchange interactions and to the usual single-ion
anisotropies, they are {\it long-ranged} and induce a {\it finite} in-plane
value of the magnetisation, although for the present planar symmetry the
Mermin-Wagner theorem, see Ref.\ \cite{Mermin}, in the presence of only
short-range interactions would rigorously predict vanishing long-range
order. Actually, the above-mentioned finite magnetisation is {\it not} in
contradiction with this theorem nor with a recent extension,
\cite{BrunoExt}, since in these papers the magnetostatic interaction is {\it
not} included.

\item Third, generically, by the magnetostatic fields and/or other
interactions, one has an effective {\it out-of-plane} anisotropy, favouring
\ben
\item an out-of-plane state of the magnetisation (this case will be
considered later), or \item an in-plane state, a case, on which we
concentrate first:\een In the above-mentioned situation, for {\it vanishing}
thickness, i.e.\ in the formal limit $t_h\to 0$, this in-plane state would
be described by a model of {\it short-range} XY-symmetry, which has a phase
transition to a low-temperature phase with diverging susceptibility, but
without long-range order for any finite $T$, the 'Kosterlitz-Thouless
transition' (KT transition), see Ref.\
\cite{Kosterlitz}. But for {\it finite} thickness, according to an equation
to be presented below,  there is always a $\vec k$-dependent {\it
long-range} effective {\it in-plane} uniaxial anisotropy {\it favouring}
e.g.\ a spin-alignment along the (arbitrary) $\pm x$-direction and {\it
disfavouring}\,\, spin fluctuations with finite in-plane $\vec k$ along the
$\pm y$-axis, i.e.\ perpendicular to the magnetisation. (E.g.\ according to
Hoffmann's well-known 'ripple theory', \cite{Hoffmann}, the so-called
'transverse ripple', with $\vec k=\pm (0,k,0)$, is disfavoured with respect
to the longitudinal one, where $\vec k=\pm (k,0,0)$, see also Ref.\
\cite{Krey0}.)

\item In fact, if the magnitude $k$ of the (in-plane) wavevector $\vec k$
times the thickness $t_h$ of the film is sufficiently low, due to long-range
magnetostatic interactions, for in-plane magnetised thin films with purely
dipolar and exchange interactions, the magnon energy $\eps(\vec k)$ is given
by the following formula, see Ref.\ \cite{Slavin}, Eqs.\ (11)--(14):
\beq\label{eqEffectiveAnisotropy}\frac{\eps(\vec
k)}{\mu_B}=\sqrt{(H_x+\frac{2A}{M_s}k^2)\cdot(H_x+4\pi
M_s+\frac{2A}{M_s}k^2)+ \frac{k\cdot t_h}{2}(4\pi
M_s)^2(\frac{k_y^2}{k^2})+...}\,,\eeq where the dots denote neglected terms
$\propto t_h^2$ or higher powers of $t_h$; i.e.\ to avoid
z-dependencies it is assumed in the following that the thickness $t_h$ of
the film, although $>0$, is nonetheless much smaller than the magnetostatic
exchange length $l_e$ defined below. Furthermore, $\mu_B$ is the Bohr
'magneton', $\mu_B\cdot\frac{2A}{M_s}\equiv D$ the spin-wave stiffness,
$M_s$ the saturation magnetisation in c.g.s.\ units, and $H_x$ a small
external field in the
$x$-direction. (As mentioned above, for $t_h=0$ this Zeeman field would
be necessary to break the $XY$-symmetry which for {\it short-range}
interactions would imply vanishing long-range order. Actually, the same
magnetisation divergence giving rise to the term $\propto k_y^2/k^2$ in
Eqn.\ (\ref{eqEffectiveAnisotropy}) (note the non-analytic behaviour on the
direction!) would also give a contribution $\propto k_yk_x/k^2$ modifying
$H_x$: this is usually neglected. However, in principle the neglection is
not allowed for small $H_x$; e.g.\ the additional terms give rise to a
'blocking phenomenon' of the magnetisation ripple, see Ref.\
\cite{Hoffmann}. But here we neglect this addition as usual, since we do not
consider that phenomenon.)

\item At the same time, due to the fact that by the strong ellipticity of
the spin-precession the {\it in-plane} amplitudes are extraordinarily large
compared with the {\it out-of plane} amplitudes, which reduces the
expectation value of the magnetisation much more than in case of circular
precession, the integrand on the r.h.s.\ of Eqn.\ (\ref{eq1}) is multiplied
by an additional '{\underbar{ellipticity factor}}' $f(\vec k)=\frac{(\eps_a(\vec
k)+\eps_b(\vec k))/2}{\sqrt{\eps_a(\vec k)\cdot\eps_b(\vec k)}}$, which is
always
$\ge 1$; i.e.\ now one would get instead of Eqn.\ (\ref{eq1}): \beq
\label{eqHeuristic}M_s(T)=M_s(0)-{\rm
const.}\times\int\limits_0^{'\pi/a'}\,{\rm d}^2k\frac{1}{\exp(\beta \eps
(\vec k))-1}\cdot f(\vec k)\,. \eeq

\ei
({\it Below, this still seemingly harmless and natural-looking equation,
which is obtained by a standard Bogoliubov-Valatin transformation, see e.g.\
p.\ 25 of \cite{Kittel}, from the bilinearized {\rm (i.e.\ approximate !)}
spin Hamiltonian of the system, will be modified by spin interactions. This
leads to a further temperature dependent phenomenological 'interaction
factor' $F_k(T)$, which is important in the Bloch region of
$d=2$-dimensional systems}.)

If one would again consider the formal limit $t_h\to 0$ of the Eqs.\
(\ref{eqEffectiveAnisotropy}) and (\ref{eqHeuristic}), then the integral on
the r.h.s.\ of Eq.\ (\ref{eqHeuristic}) would diverge, although the fraction
in front of $f(\vec k)$ is now no longer dangerous. But the 'ellipticity
factor' $f(\vec k)$ itself diverges: $f(\vec k)\to k^{-1}$ for $k\cdot
t_h\to 0$, and this ensures the logarithmic divergency of the integral in
Eqn.\ (\ref{eqHeuristic}). In fact, this divergency reflects the short-range
XY-symmetry in the limit $t_h\equiv 0$ and is in agreement with the
Mermin-Wagner theorem, see
\cite{Mermin}, since now all interactions have become short-ranged. But
since for finite, but very small thickness, one has
$\eps(\vec k)\propto (k\cdot t_h)^{1/2}\,|\frac{k_y^2}{k^2}|^{1/2}$, a
careful integration shows not only that the  integral (\ref{eqHeuristic})
converges (the dangerous part is $\propto
\int\limits_{k=0}^{'\pi/a'}\int\limits_{\phi=0}^{2\pi}\,\frac{k \,{\rm
d}k\,{\rm d}\phi_k}{(k\cdot t_h)\,(4\pi
M_s)\,\sin^2\phi_k+\frac{4A}{M_s}\,k^2}$, which is actually very subtle
since both $t_h$ and also $A/M_s$ must be different from 0), but according to
\cite{Maleev} one also gets a $T^{3/2}$-behaviour at very low temperatures.
However this is not relevant, since according to \cite{Maleev} this
$T^{3/2}$-behaviour is only observed in the lowest temperature-range, where
the dipolar coupling is all-important, e.g.\
$0\le T<T_M\cdot(\frac{t_h}{l_e})$, where $T_M=\mu_B\, 4\pi M_s/k_B$ is a
characteristic temperature for the dipolar 'shape anisotropy', while
$l_e=\sqrt{\frac{2A}{2\pi M_s^2}}$ is the already-mentioned 'magnetic
exchange length', \cite{REM1a}.

 In Permalloy, with $4\pi M_s=10^4$ Gauss and $A=1.3\,10^{-6}$ erg/cm
(remember $C_b=\mu_B\,\, 4\pi M_s$ and $D:=\mu_B\,\, (2A/M_s)$), the
temperature $T_M$ would correspond to $0.67$ K. As mentioned above, the
temperature region for {\it the Bloch behaviour is, instead, rather high}:
$T_1<T<T_2$. Estimates for the two crossover temperatures $T_1$ and $T_2$ are:

\bi\item {\underline{
$T_1\cong\sqrt{2}\cdot\mu_B\,\,4\pi M/k_B$}}; this is slightly larger than
$T_M$, but still very low compared to the temperature corresponding to the
maximal spin-wave frequency,
$T_{\rm max} \,:\cong D\cdot(\frac{\pi}{a})^2/k_B\,\,(\cong 845$ K,
with a lattice constant of $a\cong\pi\,10^{-8}$ cm, say). \item {\it But it
is known that Bloch's $T^{3/2}$-law applies to very high temperatures, e.g.\
up to {\underline{$T_2\approx T_{\rm max}/3$}}, without significant
corrections, in planar systems of the above-mentioned kind,
\cite{Gradmann1,Koebler}. And it should be stressed again at this place that
this wide temperature region (the 'Bloch region') is still  below the
thermal critical region,
\cite{REM4}}.\ei

In contrast, the relation $\eps(\vec k)\propto (k\cdot
t_h)^{1/2}\,|\frac{k_y^2}{k^2}|^{1/2}$ is only valid for $k\lsim k_c$, where
$k_c:=t_h/l_e^2$ is a small 'crossover value': For larger
wavenumbers the exchange interaction starts to dominate, i.e.\ $f(\vec k)\to
1$, and \beq M_s(T)\cong M_s(\tilde T)-{\rm const.}\times
\int\limits_{k_c}^{'\pi /a'}\, 2\pi k {\rm d}k\frac{1}{\exp(\beta D\cdot
k^2)-1}\,, \eeq where $k_B\tilde T\,:=\,\eps (k_c)$. By numerical
evaluation (see below) this would inevitably lead to some kind of
approximate 'D\"oring behaviour', \cite{Doering}, i.e.\ to
$M_s(T)=M_s(\tilde T) -{\rm const.'''}\times T\cdot\ln \frac{T}{\tilde
T}$, where $\tilde T$ can be considered as a temperature unit entering
e.g.\ by quantum effects enforcing long-range order (In fact it turns
out below that $\tilde T$ is identical with $T_1$). As far as the author
knows, the 'D\"oring behaviour' has never been observed, although, in
the temperature range considered, it was also suggested by Maleev,
\cite{Maleev}, in a paper preceding that of Bruno, \cite{Bruno}.

\bi\item
{\it Instead, to get the observed $T^{3/2}$-behaviour (see
\cite{Gradmann1,Koebler}) also in the 'D\"oring temperature range',
i.e.\ also for $T\rsim T_1$, one should note that, as soon as the
wave-number $k$ is large enough, there is no longer any in-plane
uniaxial anisotropy felt by the magnons}.\ei

This means that it is natural to assume, as in the $d=2$-dimensional XY
model, that there exists a strong enhancement of the {\it destructive
character} of these magnetic excitations, 'destructive' with respect to the
magnetic long range order, leading to an additional temperature-dependent
factor $F_k(T)$ which represents the interaction of the magnetic
quasiparticle excitations. In the planar XY model, this additional
'destructive factor' acts on all $k$-scales ($T$-scales) until
$k=0$ ($T=0$); therefore, the magnetic long range order is completely
destroyed in the XY-model. Physically, this happens through the dynamic
formation of magnetic vortex excitations: In the low-temperature phase of
the KT transition these vortices are strongly bound in pairs of opposite
topological charge (opposite chirality), and the typical distance of the
vortex centers of such 'vortex pairs', $\langle l\rangle_T$, remains finite,
if the critical temperature is attained: At this temperature there is {\it
vortex unbinding} at $\langle l\rangle_{T_c}=l_V$.

\bi\item{\it
As a consequence, spinwaves are not influenced by the vortex formation, if
their wavelength $\lambda =2\pi /k$ is {\underline{much larger}} than $l_V$,
whereas  spatial spin-spin correlations with distances {\underline{much
smaller}} than $l_V$ are wiped out by the vortices.
}\ei

 Similarly, we assume that in the Bloch region there is for $k>k_c$ a
'reduction factor' $F_k(T)$ representing the enhanced reduction of magnetic
order by the interactions between magnetic excitations of wavenumber $\vec
k$. In the present case, this reduction factor is not as fatal as in the
$d=2$-XY model, since wavenumbers smaller than $k_c$ are not involved. But
one has again  some kind of Pippard heuristics: Namely,
along the lines of the original arguments of Kosterlitz, \cite{Kosterlitz2},
we assume that $F_k(T)\propto \langle k\rangle_T\cdot l_V$, where $\langle
...\rangle_T$ denotes a thermal average, so that the reduction increases
with increasing $T$, i.e.\ increasing $\langle k\rangle_T$. In fact, according
to the (by now already known) Pippard argument, $\langle k\rangle_T$ can be
replaced by
$(k_BT/D)^{1/2}$. Thus again $M_s(T)=M_s(\tilde T)- {\rm const.}\times
T\times T^{1/2}$, where on the r.h.s.\ the factor $T$ represents the
'equipartition energy', while (for
$d=2$) the
factor $T^{1/2}$ arises from the 'reduction factor': This factor represents
the strong reduction of long-range order by spin interaction effects, which
are strongly 'relevant' in
$d=2$. (In $d=3$ dimensions the 'reduction by interaction' is apparently not
so important in the Bloch region. There, instead, the magnetic moment of
whole columns of sites with identical planar coordinates (x,y) is reduced by
the additional fluctuations corresponding to dynamic noncollinearities of
the spins in the
$z$-direction. This effect for
$d=3$ is already described with noninteracting magnons by our previous
factor $PhsR(T)$. Thus, $d=2$ is much more dangerous to interactions than
$d=3$,  in agreement with the fact that the 3-dimensional $XY$-model is
'harmless' compared with the $d=2$-dimensional $XY$-problem.)

There is another  argument  concerning $F_k(T)\propto T^{1/2}$: Since the
typical thermal energy of an entity representing one degree of freedom is
$k_BT$, and since the interaction should be dominated by two-body
scattering, one guesses that in systems with strong magnon-magnon
interactions the quasiparticle numbers $\langle n_k\rangle_T=\{\exp(\beta
\eps(\vec k)-1\}^{-1}$ should be multiplied in the present context by
$(k_BT)^{1/2}$\,: This concerns the first particle of an interacting pair. A
similar interaction factor
$(k_BT)^{1/2}$ should come from the second particle; i.e.\ together the
two interacting particles might be considered as an entity involving the
genuine thermal energy $k_BT=[(k_BT)^{1/2}]^2$. This is actually again some
kind of Pippard argument; it will be repeated later, when the transition
from
$s=1/2$ to $s=1$ is considered in section 8.

\section{Quantum effects and crossover-flow for d=2}
It is important to understand what is going on: Therefore in the following
we sketch a flow-line scenario for an in-plane magnetised thin film.
Generally, we consider in the following a cartesian coordinate system where
on the abscissa axis ('$x$-axis') we plot the effective out-of-plane
anisotropy energy
$C_b\,\,\,(e.g.\,\,C_b=\mu_B 4\pi M_s$), which favours an
in-plane magnetisation and is at the same time proportional to an effective
uniaxial anisotropy energy (e.g.\
$\eps(\vec k)\cong\mu_B 4\pi M_s\, [(k\cdot
t_h/2)\sin^2\varphi_k]^{1/2}$) favouring  an (arbitrary) $\pm x$-direction
as in-plane magnetisation direction.

On the ordinate ('$y$'-axis) of our plot the reduced Kelvin temperature
$k_BT$ is presented. The origin of the cartesian coordinate system is at
$(C_g=0,T=0)$.

On the third axis ('$z$-axis) the magnetisation $M_s(C_b,T)$ is presented,
or any other thermal expectation value depending on $C_b$ and $T$.

Along the $x$-axis for $T=0$,  one  must consider quantum fluctuations
(''zero-point fluctuations''), which by the ellipticity of the spin
precession reduce the magnetisation from the equivalent of the spin quantum
number $s$ to an effective value, which is smaller:

 $\langle \vec k|\hat S^z_{\vec l}|\vec k\rangle=s\quad\quad\to\quad\quad\langle \vec
k|\hat S^z_{\vec l}|\vec k\rangle = s-|\gamma_2(\vec k)|^2$.

 Here $|\vec k\rangle :=\sum\limits_{\vec r_l}\,\frac{\hat S_{\vec l}^x-{\rm
i}\,\hat S_{\vec l}^y}{\sqrt{2sN}}\,\exp( {\rm i}\vec k\cdot\vec r_{\vec
l\,}\,\,)\,|GS\rangle$, where $|GS\rangle$ is the (non-trivial~!) ground
state of the system. The vectors $\vec l$ count the $N$ sites $\vec r_{\vec
l}$ of the lattice considered, the $\hat S^x_{\vec l}$ etc.\ are the spin
operators.

 By the Bogoliubov-Valatin transformation, see e.g.\ page 25 in Ref.\
\cite{Kittel}, the ground-state $|GS\rangle $ is not identical with the
'trivial state' $|0\rangle$, where all spins are aligned in the preferred
direction (i.e.\ $|0\rangle \hat = |m_l=s\rangle$ $\forall l$, i.e.\ for all
sites of the lattice). Instead, $|GS\rangle$ is obtained from $|0\rangle$ by
a straightforward calculation, which hides the 'ellipticy' by a mathematical
formalism described in detail in the above-mentioned 'text-book reference'.
In this way one derives the following identity involving the
'ellipticity factor' $f_k=\frac{(\eps_a(\vec k)+\eps_b(\vec
k))/2}{\sqrt{\eps_a(\vec k)\cdot\eps_b(\vec k)}}$ of Eqn.\
(\ref{eqHeuristic}) and the 'zero-point spin reduction' $|\gamma_2(\vec
k)|^2$: 
\beq |\gamma_2(\vec k)|^2=\frac{1}{2}\cdot (f_k-1)\equiv\eeq $$
\frac{1}{2}(\frac{(2A/M_s)k^2+2\pi M_s+\pi M_s\cdot (k\cdot
t_h)\cdot\sin^2\varphi_{\vec k}}{\sqrt{((2A/M_s)k^2+4\pi M_s)\cdot
((2A/M_s)k^2+2\pi M_s\cdot (k\cdot t_h)\cdot\sin^2\varphi_{\vec
k})}}-1)\,.$$

 In Fig.\ 1 the zero-point spin reduction $|\gamma_2(k)|^2$ is plotted as a
function of $k$ for $\sin^2\varphi_{\vec k}=1$ with Permalloy material
parameters (i.e.\
$4\pi M_s=10^4$ Gauss, $A=1.3\times 10^{-6}$ erg/cm) and a thickness of
$t_h=4$ nm, as in the magnetic nanostructures studied by
H\"ollinger {\it et al.}, \cite{REM1a}.

As can be seen from Fig.\ 1, $|\gamma_2((0,k,0))|^2$ is less than $10 \,\,\%$
for $k\rsim 0.1$ (nm)$^{-1}$; for these $k$-values the linearized
Bogoliubov-Valatin theory makes sense. Whereas for $k$-values below that
value, corresponding to almost macroscopic wavelengths, there is a strong
increase of the ellipticity so that for these small wavenumbers the
linearization would be no longer meaningful.

But for $t_h\equiv 0$ one sees that  for $k\cdot l_e\ll 1$, where
$l_e:=\sqrt{\frac{2A}{4\pi M_s^2}}=\sqrt{\frac{D}{C_b}}$ is the
above-mentioned exchange-length ($\approx 5.7$ nm for Permalloy), a
divergency develops for $k\to 0$:\,\, $|\gamma_2(\vec k)|^2\approx
\frac{1}{2}\cdot f_{\vec k}=1/(4k\cdot l_e)\gg 1$. This divergency
has already been mentioned above, and in the present context for $T=0$ it
means that for small enough values of $k\cdot l_e$ spatially periodic spin
fluctuations of this small wavenumber $k$ cannot be treated any longer in
the Bogoliubov-Valatin theory (i.e.\ by linearization), since e.g.\ the
effective spin-quantum number $s^{\rm eff}=s-|\gamma_2(\vec k)|^2$ becomes
negative for these $k$. This is true although the integral $\int {\rm
d}^2k\,|\gamma_2(\vec k)|^2$ is finite (whereas, as already mentioned above,
for $t_h\equiv 0$ and finite temperatures, long-range order would not at all
exist, i.e.\ $\int {\rm d}^2k\,\frac{|2\cdot\gamma_2(\vec k)|^2+1}{\exp
\beta\eps(\vec k)-1}\equiv\infty$). In contrast, for $t_h>0$, $|\gamma_2(\vec
k)|^2$ would diverge in a similar way, namely as $\approx
0.25/\sqrt{(k\cdot t_h/2)\,\sin^2\varphi_{\vec k}+k^2\cdot l^2_e}$, where
the first term depends essentially only on the thickness and not on the
other parameters, but long-range order at finite $T$ exists in this case
nonetheless for all values of $C_b$. So the convergence of the integral for
$M_s(T)$ and the behaviour of the zero-point reduction $|\gamma_2(\vec
k)|^2$ for small $k$ are not simply related.

{\it But as a consequence of the preceding paragraphs, due to the magnetic
dipole interaction there is automatically a $k$-dependent effective
in-plane uniaxial anisotropy favouring the axis parallel to the
magnetisation. This $k$-dependent effective anisotropy can be roughly
obtained by replacing the effective field $2\pi M_s\cdot k\cdot
t_h\cdot\sin^2\varphi_k$ by its angular average
$2\pi M_s\cdot k\cdot t_h\cdot 0.5$. Here the typical long-wavelength scale,
e.g.\ the typical wavelength in the observation of magnetisation ripple
phenomena, see Ref.\ \cite{Hoffmann}, is given by the characteristic
wavenumber
$k=1/l_e$. So we associate to the magnetostatic phenomena in thin films
a k-dependent effective in-plane uniaxial anisotropy of the order of
$C_u(k):=\mu_B\cdot \pi M_s\cdot k\cdot t_h$, i.e.\ we write $C_u(k)=\tilde
C_u\cdot k$, with $\tilde C_u:= \mu_B\cdot \pi M_s\cdot t_h$. Here $k\lsim
1/l_e$.}

In Fig.\ 2, again for the already mentioned Permalloy parameters, we present
results for the 'naive magnetisation decay' $M_s(T)$, i.e.\ calculated
according from the formula (\ref{eqHeuristic}) for $d=2$ with the
'ellipticity factor' $f_k$, but {\it without} the 'interaction factor'
$F_k(T)$. By comparison with similar results for $d=3$, where the $T^{3/2}$
behaviour is quite evident, we get for $d=2$ the already mentioned 'D\"oring
behaviour', i.e.\ the result looks like $(M_s(T)-M_s(T_0))\propto T\cdot \ln
\frac{T}{T_0}$, where $T_0$ is a temperature representing an effective
'cutoff' at low k (see the figure caption of Fig.\ 2). 

\noindent
{\it Since the experimental observations deviate apparently from the D\"oring
behaviour in a significant way, the additional 'interaction factor',
mentioned above, seems really necessary}.

On the horizontal coordinate axis of our plot one has thus an 'ordered line
segment' ranging from the above-mentioned value $C_u$, calculated e.g.\ for
$k\equiv 1/l_e$ or any other relevant characteristic reciprocal length of
the system, up to infinity, whereas the dangerous 'infrared ellipticity
fluctuations' corresponding to smaller wavenumbers are 'gapped away' at very
low temperatures, since the amplitudes behave $\propto
\exp(-\frac{C_u}{k_BT})$.

From the value on the x-axis corresponding to the above-mentioned
standard case (e.g.\ Permalloy) on the 'ordered line segment' a flow line
emerges, which is well-defined if along the third axis ('$z$'-axis) a
quantity like $M_s(C_b,T)$ is plotted, and which flows to the thermal
critical point.

This line begins at $T=0$, i.e.\ in the {\it quantum regime}, with a
positive vertical slope, but soon it may turn to the left, since with
increasing $T$ the typical wavenumbers increase, the ellipticity decreases,
and the exchange interactions become important. So the flow line in this
region may resemble (details are not needed in the following !) a flat
elliptical segment running almost parallel to the x-axis in the direction of
the y-axis, until it extrapolates, almost horizontally, to the
above-mentioned small crossover temperature $T_1\,\,(\cong\sqrt{2}\,T_M$,
see below). This should signal the crossover from the {\it linear} rsp.\
{\it square root} behaviour
$\eps(\vec k)\cong\sqrt{C_b\cdot Dk^2+\tilde C_u\cdot k}$ of the dispersion
for $k\cdot l_e\lsim 1$ to {\it parabolic} behaviour, $\eps(\vec k)\cong D\cdot
k^2$ for $k\cdot l_e\rsim 1$. Therefore one roughly expects $k_BT_1\cong
\sqrt{2}\, C_b$, see e.g.\ Eqn.\ (\ref{eqEffectiveAnisotropy}) for
$t_h=0$: It is, in fact, visuable from Fig.\ 3, that there is a rather sharp
crossover of this kind at $k_c:=\pi t_h/l_e^2\cong 0.1$ $(nm)^{-1}$. At
shorter wavelengths (larger $k$), the exchange interaction dominates and the
interaction energy is $\approx k_BT_1$ or larger.

It is not essential (and not at all universal) how this 'crossover point' is
reached, i.e.\ from which ground state at $T=0$ (as we have seen the ground
states will be nontrivial, i.e.\ not given by the Bogoliubov-Valatin
transformation at length scales $k\cdot l_e\ll 1$). In contrast, in the
Bloch region $T_1<T<T_2$, where the dispersion of the magnons, whether
interacting or not, is dominated exclusively by the exchange interaction,
the behaviour is essentially universal, e.g.\ the values of the
anisotropies, including the effects of the magnetostatic interactions,
become irrelevant.

Analogously, at the second axis ('y'-axis) one has  the following scenario:
long-range magnetic order up to the thermal critical temperature (Curie
temperature) $T_c$, which is of the order of the maximal magnon energy,
$T_{\rm max} :\cong\, D\cdot (\frac{\pi}{a})^2/k_B$, or typically even
larger, \cite{REM4}. From this Curie fixed point, i.e.\ in the {\it
thermal regime},  our (backward) flow line emerges starting at first
horizontally, but soon turning downwards, until it flows back from
$T_2\, :\cong T_{\rm max}/3$ towards  $T_1\,\,$ essentially according to
Bloch's $T^{3/2}$-law, thereby reaching the above-mentioned crossover point,
where $k_BT_1\approx \sqrt{2}\,C_b$, as already noted. From there the
(backward) flow returns to the starting-point (corresponding e.g.\ to
ultrathin Permalloy films with finite $t_h\ll l_e$) on the 'ordered line
segment' of the $x$-axis, where the linearized (i.e.\ Bogoliubov-Valatin)
transformation should be valid until the very small effective in-plane
uniaxial anisotropy of the order of $C_u\equiv C_b\cdot\frac{t_h}{l_e}$: But
this last part of the flow is non-universal, i.e.\ $C_u$ can also be
replaced by a quantity not depending on $C_b$.

{\it Now, what counts in the Bloch behaviour between $T_1$ and $T_2\cong
\frac{1}{3}\,T_{\rm max}$, are the thermal energy factor $\propto T$ and the
effective interaction radius (in $\vec k$-space) $\propto T^{1/2}$
corresponding to only the segment of the flow line between $~T_1$ and $T_2$
(in this region the quasi-classical 'poor man's scaling' considerations
apply)}.

\section{Two-dimensional systems with positive out-of-plane anisotropy}
For two-dimensional systems with a {\it positive} uniaxial {\underline{\it
perpendicular}} anisotropy, i.e.\ favouring (in contrast to the previous
case) an out-of-plane orientation, there is no ellipticity of the spin
precession around the $z$-direction, and the temperature dependence of the
magnetisation follows in a naive formulation analogously to Eqn.\
(\ref{eqHeuristic}), namely $M_s(T)=M_s(0)-{\rm const.}\cdot \pi\times
\int\limits_{k=0}^{\pi /a}\,\frac{2k{\rm d}k}{\exp\beta\eps(k)-1}$, with
$\eps(k)=C_b^{\rm out}\,+\,D\cdot k^2$, where $C^{\rm out}:=-C_b-4\pi M_s
\,> 0$, whereas in-plane magnetostatic fields are now neglegible. 

The above-mentioned integral can be evaluated analytically, yielding a
variant of the 'D\"oring behaviour', namely \beq M_s(T)\equiv M_s(0)-{\rm
const.}\cdot\pi\times \left (\frac{k_BT}{D}\right )\cdot \left
(-\frac{D\cdot (\pi /a )^2}{k_BT} +\ln\frac{e^{x_{\rm max}}
-1}{e^{x_{\rm min}}-1}\right )\,, \eeq with $\eps_{\rm max} :=\beta\cdot
(C_b^{\rm out}+D\cdot (\frac{\pi}{a})^2 )$ and  $\eps_{\rm min}
:=\beta\cdot C_b^{\rm out}$.

Again, if in accordance with \cite{Koebler} a $T^{3/2}$-behaviour would be
observed, an additional $T^{1/2}$-factor would be necessary.

In the particular case of a 'reorientation transition', corresponding to a
situation, where the effective uniaxial anisotropy would be zero,
$C_b^{\rm out}\equiv 0$, for {\it short-range interactions} the original
Mermin-Wagner theorem would become applicable and the magnetic order would
vanish: Instead, again by the (long-range) dipolar interactions, one
observes a nontrivial domain structure.

 At this place it should be be mentioned that the 'reorientation
transition',  and  in particular spinwave excitations (including the dipolar
interactions and surface anisotropies) in systems with a
finite number of non-equivalent magnetic planes with different tilt angles
of the magnetization direction in these planes, have been carefully studied
in a series of papers by Erickson and Mills,
\cite{Mills1,Mills2,Mills3}.

\section{d=1:  $T^{5/2}$-behaviour ?}
Here we consider a homogeneous ferromagnetic wire stretching from
$x=-\infty$ to $x=+\infty$ along the $x$-axis, with constant circular cross
section, magnetised longitudinally, i.e.\ in the x-direction. Now the
dipolar (effective) anisotropy acts with equal strength, $C_b=\mu_B\cdot
2\pi M_s$, along the $y$-axis and the $z$-axis, so that, by applying the
'naive approach' of Eqn.\ (\ref{eqHeuristic}) again, we get (almost as
before) $M_s(T)=M_s(0)-{\rm const.}\times\int\limits_0^{ \pi /a }\,\frac{{\rm
d}k}{\exp \beta\cdot
\eps(\vec k) \,-1\,}$, with $\eps_k=\mu_B\cdot 2\pi M_s +D\cdot k^2$. For
temperatures allowing the approximation $\exp(\beta\eps(\vec
k))-1\cong\beta\eps(\vec k)$ this can be directly evaluated and yields
$M_s(T)=M_s(0)-{\rm const.}\times T\times{\pi /2}{\sqrt{\mu_B\cdot 4\pi
A}}$, whereas for general temperatures one gets the numerical behaviour
presented in Fig.\ 4, i.e.\ also a  behaviour $\propto T$.

That this behaviour cannot be valid to arbitrary low temperatures follows
from the Maxwellian relation $\partial S/\partial H=\partial M/\partial T$
and Nernst's 'Third Fundamental Theorem' of thermodynamics, according to
which
$\partial S/\partial H\to 0$ for $T\to 0$.

In fact, experimentally one has obtained instead, see \cite{Koebler}, a pronounced
$T^{5/2}$-dependence, in agrrement with the apparently universal
classification of U.\ K\"obler for half-integer $s$ and one-dimensional
systems,
\cite{Koebler}, and astonishingly also with the experiments by U.\ Gradmann
and coworkers, \cite{Gradmann2}, on $Fe$ monolayers on W(110) (covered by
Ag). Also the films with the different $T^{3/2}$-behaviour studied in
\cite{Gradmann1} are planar $Fe$ films on W(110), but thicker (from 3 to 21
monolayers), and whereas (as already mentioned above) in \cite{Gradmann1}
one has the $T^{3/2}$ behaviour discussed in the two preceding sections, for
the films of
\cite{Gradmann2} one has instead the $T^{5/2}$ behaviour, which is still
more astonishing. Not only for this system, but also for $d=1$ this is
unexpected and suggests that here the 'interaction damage' factor $F_k(T)$
is even stronger, namely
$\propto T^{3/2}$ instead of $\propto T^{1/2}$. Presently, this is not
at all understood, or only very qualitatively: One knows that in one
dimension the critical thermal fluctuations destroy long range order at
finite $T$ even for Ising models, whereas in $d=2$ the fluctuations are
substantial for $XY$-models, but not for Ising models. So it is
not quite surprising that already for the Bloch region the interaction
factor should be much stronger for $d=1$ than for $d=2$.

Topological excitations may play an essential role also in this case, namely
the so-called Bloch-point excitations, which are studied in a recent paper
of Thiaville {\it et.\ al.}, \cite{Thiaville}. These are very 'floppy'; they
consist topologically of opposite spin directions along a certain axis,
e.g.\ $\vec S\propto \pm
\hat z$ for $z > 0$ rsp.\
$z<0$, but in the $xy$-plane there can be a vortex. In this case, the
'damage factor' $F_k(T)$ corresponding to  vortex interactions exists for
all three orthogonal planes, the $xy$-plane, the $yz$-plane and the
$zx$-plane.

 In any case, the $T^{5/2}$-behaviour in the Bloch region does not
contradict the fact that ultimately, i.e.\ after a further crossover to the
genuine thermal critical region, there can exist not even Ising critical
order in the wire, such that finally $T_c\to 0$. For Gradmann's monolayer
films, \cite{Gradmann2}, instead, after the {\it pseudo-one-dimensional}
(non-Ising) behaviour in the Bloch region, the system should return to the
genuine two-dimensional critical (Ising-) behaviour near $T_c$.

So this is quite complicated and subtle, as usual if one is dealing with
dipolar interactions.


\section{$T^{2}$ crossover-scaling  for itinerant magnets in d=3} 
The arguments of the following section look especially simple (although this
may be deceptive): We consider a
crystalline metallic ferromagnet in d=3 dimensions. The system has separate
energy bands $\{E_\uparrow (\vec k)\}$ rsp.\ $\{E_\downarrow(\vec k)\}$, and
at $T=0$ the 'Stoner excitations', i.e.\ electron-hole excitations with spin
flip, where e.g.\ an electron with energy 'immediately below the Fermi
energy $E_F$', with initial-state wave-vector $\vec k$ and spin
$\uparrow$, is moved to a state 'immediately above $E_F$', with a different
final-state wave-vector $\vec k^{\,'}$ and spin $\downarrow$. Now at finite
$T$, the expression 'immediately above or below $E_F$' means: 'within an
interval of width $\approx k_BT$ around $E_F$. Every such excitation reduces
the magnetic moment of the sample by two Bohr magnetons; i.e.\ in the
above-mentioned 'heuristical formula' one has a 'phase-space factor'
$\propto (k_BT)^2$, where one of the two powers of $k_BT$ stands for the
effective phase-space of the {\it initial states}, while the other one
represents the {\it final states}. There is no 'energy factor', since all
this happens in the above-mentioned interval immediately near
$E_F$.

Similar arguments also apply to itinerant {\it antiferromagnets},  in
accordance with the observations of U.\ K\"obler, \cite{Koebler}, who always
observed the same Bloch exponent for the magnetisation of ferromagnets rsp.\
the sublattice magnetisation of antiferromagnets (As a consequence the Bloch
scaling is probably not directly related to quantum-criticality, since the
quantum critical behaviour of ferromagnetic and antiferromagnetic systems
is different). In contrast, for semiconducting systems as $EuO$ and $EuS$
the magnetism should better be described by a (non-itinerant) $d=3$
Heisenberg model with
$s=7/2$, and so one expects the $T^{3/2}$-behaviour for $d=3$. But an
'interaction factor' of $T^{1/2}$ would transform this into the observed
$T^2$ behaviour, in accordance with \cite{Koebler}. Yet such a behaviour
would definitely go beyond the Heisenberg model, since within that model
there is the rigorous analysis of Dyson, Ref.\ \cite{Dyson}, for arbitry
fixed spin quantum number
$s$, see Ref.\ \cite{Dyson}, predicting a more gradual decrease:
$M_s(T)=M_s(0)
-c_1\cdot T^{3/2}-c_2\cdot T^{5/2}-c_3\cdot T^{7/2}-c_4\cdot
T^4-{\cal{O}}(T^{9/2})$. Note that the first term with {\it integer}
exponent is $\propto T^4$, and not $\propto T^3$; the integer term results
mainly from the decrease of the magnon stiffness, which according to
rigorous theories 'a la Dyson' should be $\propto T^{5/2}$, whereas more
phenomenological theories, and according to K\"obler, Ref.\ \cite{Koebler},
also the experiments, suggest a decrease following an effective behaviour as
$D(T)=D(0)-{\rm const.}\cdot T^{3/2}$.

 Because of the large value of $s=7/2$ for Eu$^{2+}$, the 'non-Dysonian'
behaviour, and the 'extra' interaction factor $\propto T^{1/2}$, look not
completely unreasonable. And one should note again that this behaviour goes
definitely beyond the (underlying) Heisenberg model, which always assumes
fixed
$s$, in contrast to the Pippard arguments from the previous section and
those discussed below in connection with the problem of the difference
between integer and half-integer $s$. And one should also note, concerning
the case of metallic magnetism, that the Stoner mechanism (which is actually
a two-particle mechanism) looks completely different from anything
formulated in the Heisenberg model. In contrast, the {\it collective} magnon
excitations, which are also present in itinerant magnets and which have the
same quadratic dispersion as before, seem to give a neglegible contribution
to the behaviour of $M_s(T)$ in the temperature range considered, although
at very low temperatures a
$T^{3/2}$-contribution would always be larger than a $T^2$-one. So this
means again that one should remain above a certain crossover-temperature, in
this time that one to the dominance of the $T^{3/2}$ behaviour for $d=3$.

\section{Integer $s$}
The arguments in the previous sections have all considered half-integer
values of $s$, e.g.\ $s=1/2$ and $s=7/2$. For integer values, e.g.\ $s=1$, one must
consider the additional interaction energy, $\sim k_BT$, which is necessary
to break e.g.\ a $s=1$ triplet state of two $s=1/2$ spins into two separate
$s=1/2$ entities. This happens at sufficiently high temperatures at the
upper range of the Bloch region, i.e.\ below the genuine critical temperature
region.

So for a single entity the {\it square root} of that energy comes into play
as an additional energy factor so that K\"obler's effective Bloch exponents
of $\eps_B=3/2$ (rsp.\ $\eps_B=5/2$) for spin systems with half-integer $s$ and
(effective) Bloch-dimensionality $d=2$ (rsp.\
$d=1$) should change for integer $s$ to $\eps_B =4/2$ (rsp.\
$\eps_B =6/2$), apparently as observed, see \cite{Koebler}. For $d=3$ things
are more complicated: according to the tentative arguments of the previous
section, the Bloch exponent $\eps_B=2$ for $d=3$ and half-integer $s$ is given
by a product of three terms, which can be written symbolically as
$(k_BT)^1_{\rm initial\,\,state}\times(k_BT)^1_{\rm
final\,\,state}\times(k_BT)^0$, where the third term represents the constant
energy-factor. However for integer $s$, due to the above-mentioned
interaction effects, the first two factors should each acquire an additional
power of $k_BT$, while the third factor should acquire an additional
'square-root power' $(k_BT)^{1/2}$. In this way, starting with the
$T^2$-behaviour for half-integer $s$ one would expect a
$T^{9/2}$-behaviour for integer $s$, as observed, \cite{Koebler}. Whether
this  is more than {\it heuristics}, has to be shown, similarly to the
previous sections, but it should be noted that the pronounced crossover from
a $T^2$ low-temperature behaviour to a $T^{9/2}$ behaviour at higher
temperatures, which U.\ K\"obler observes for $NiO$ at $T=100$ K rsp.\ for
$Fe$ at 628 K, seems to be in favour of the present scenario \footnote{At
several places in the observations of U.\ K\"obler, see \cite{Koebler}, also
a crossover from a $T^{3/2}$-behaviour 'at low temperatures', e.g.\ for
$T\lsim 200$ K, to a $T^2$-behaviour at a considerable range of higher
temperatures is observed.}.

\section{Conclusions}{ As a consequence of the preceding sections,
concerning the temperature range considered, it should be noted that the
arguments of the present paper, supported by quantitative calculations
especially in sections 2 and 3, do neither concentrate on the magnetostatic
interactions (although these are very important, as seen in those chapters)
nor stress the immediate vicinity of the quantum region at
$T=0$ nor that of the thermal critical region near $T_c$, but rather
concentrate on the (wide) 'Bloch temperature range' between the (rather low)
crossover temperature
$T_1$\,\,\,\, {\rm (}$\cong\sqrt{2}\,\,C_b/k_B$) {\rm and a (much higher) crossover
} temperature of the order of magnitude of $T_2\cong\frac{1}{3}\,D\cdot
(\pi/a)^2/k_B$.

 {\it So  quantum behaviour {\underline{\rm is}} significant, but at the
same time, in the Bloch region, it only plays a secondary role, apparently
in contrast to {\rm interactions}, which appear to be particularly important for
$d= 2$, where they are known to destroy the long-range order for the
Kosterlitz-Thouless phase transition. In higher dimensions also the main
implicit assumption of the Heisenberg model, that the spin quantum-number
$s$ is invariably constant, becomes questionable through the interactions
already in the Bloch region.}

All statements in the previous sections, except from section 8, apply to
half-integer $s$ in K\"obler's classification, \cite{Koebler}. Already at
this place, i.e.\ for half-integer s, the role of the interactions seems
crucial, particularly for $d=2$ and $d=1$. Additional complications for
integer $s$ (outside the region of overwhelming importance of the thermal
fluctuations) have been discussed in section 8.

 A more thorough analysis of the questions left open, which mainly concern
the role of interactions, is shifted to future work.

 \section*{Acknowledgements} The author acknowledges stimulating
 communications from K.\ Binder, P. Bruno (who pointed out an error in an
 earlier version of this paper), U.\ Gradmann, U.\ K\"obler and S.\
 Krompiewski, and fruitful discussions with C.\ Back, G.\ Bayreuther, R.\
 H\"ollinger, A.\ Killinger, W.\ Kipferl, and J.  Siewert.

\vglue 0.3 truecm
\noindent

\newpage

\centerline{\underline{\bf Figures and Figure Captions}}
\vglue 1.0 truecm
\hglue 2.4 truecm
\epsfxsize=7cm
\epsfbox{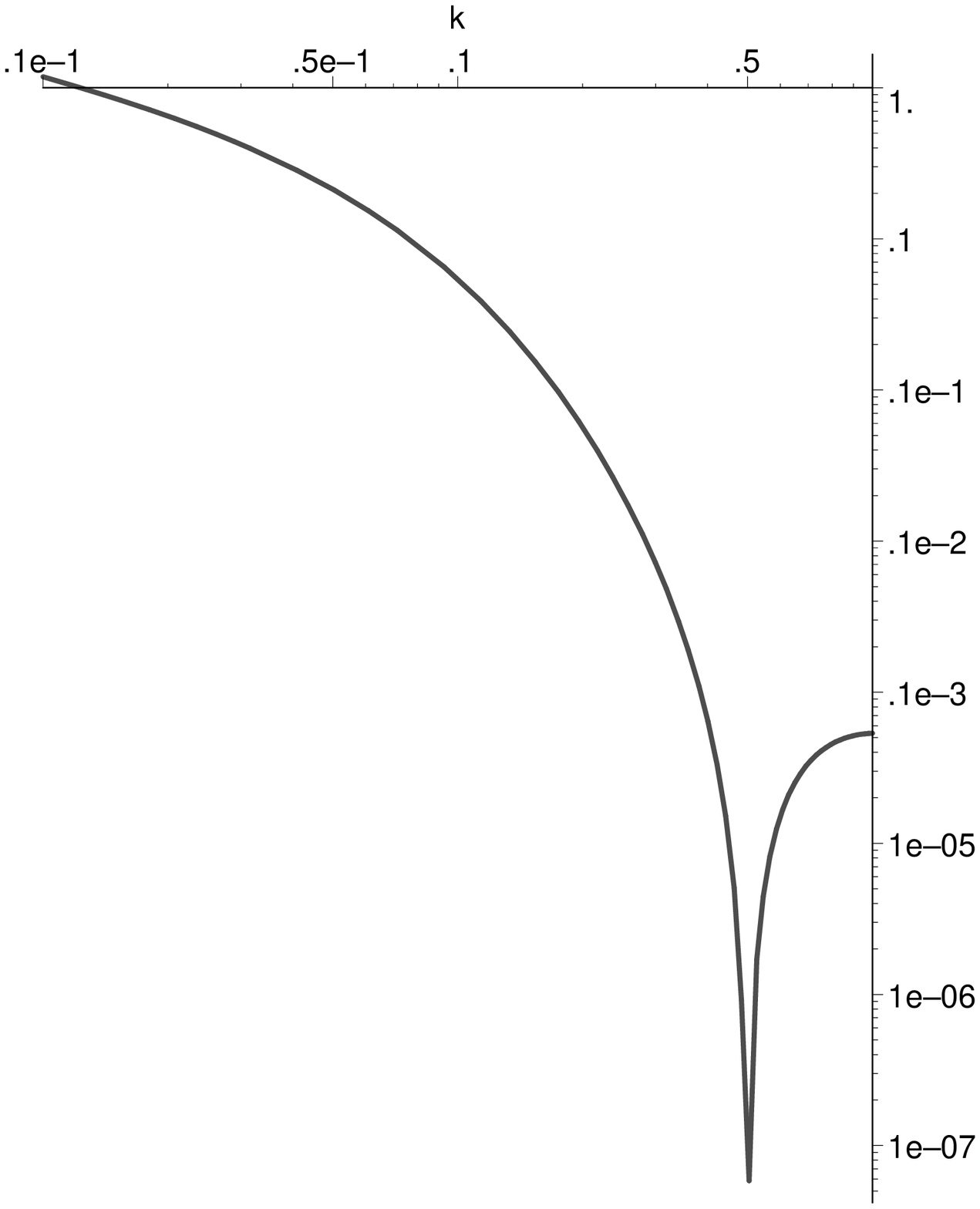}
\vglue 1.0 truecm

\noindent{\bf Fig.1:}
{{{The zero-point spin reduction $|\gamma_2(\vec k)|^2$ (see the text,
especially Eqn.\ (\ref{eqEffectiveAnisotropy})) is presented over the
wavevector $\vec k= (0,k,0)$ , where $k$ is measured in units of
$(nm)^{-1}$, for an ultrathin Permalloy film of thickness $t_h=4$ $nm$, in a
double-logarithmic plot with material parameters $4\pi M_s=10^4$ Gauss,
where $M_s$ is the cgs-magnetisation, and $A=1.3\times 10^{-6}$ erg/cm the
spin-wave stiffness.

The characteristic magnetostatic exchange length $l_e=(2A/(4\pi M_s))^{1/2}$
is 5.7 nm, such that the characteristic value $k_c:=t_h/(2l_e^2)$ is
$\approx 0.06$ (nm)$^{-1}$; $|\gamma_2(\vec k)|^2$ vanishes at $k=\pi /l_e$ i.e.\
at $\approx 0.5$ $(nm)^{-1}$. For $k\rsim 0.1$ (nm)$^{-1}$ it is less than
$10 \,\,\% $. For these $k$-values the linearized Bogoliubov-Valatin theory
makes sense. Whereas for $k$-values below that value, corresponding to
almost macroscopic wavelengths, there is a strong increase of the
ellipticity so that for these small wavenumbers the linearization would be
no longer meaningful.

}}}

\newpage


\hglue 0.5 truecm\fbox{\hglue 0.35 truecm\epsfxsize=4cm
\epsfbox{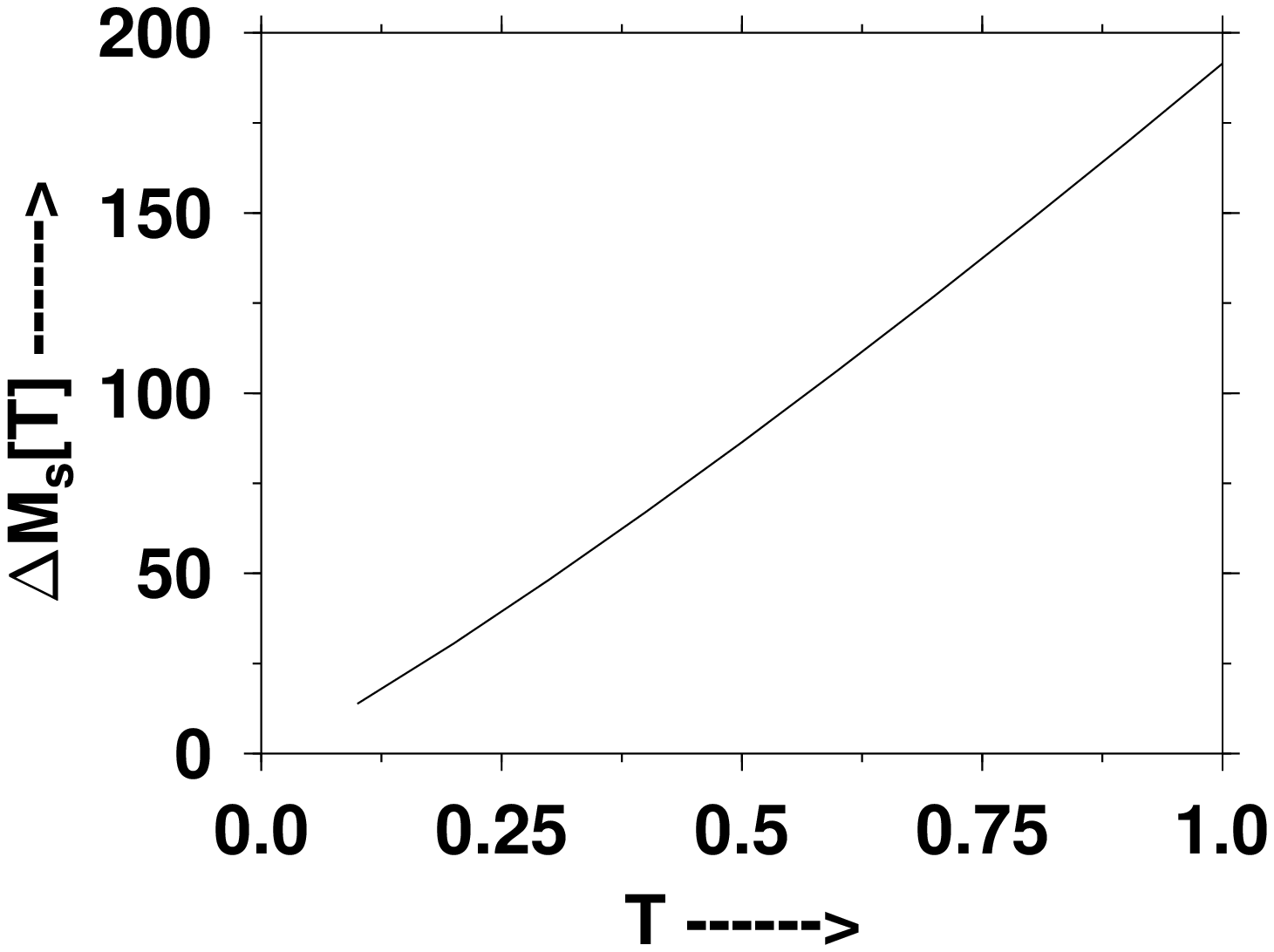}
\hglue 0.8 truecm
\epsfxsize=4cm
\epsfbox{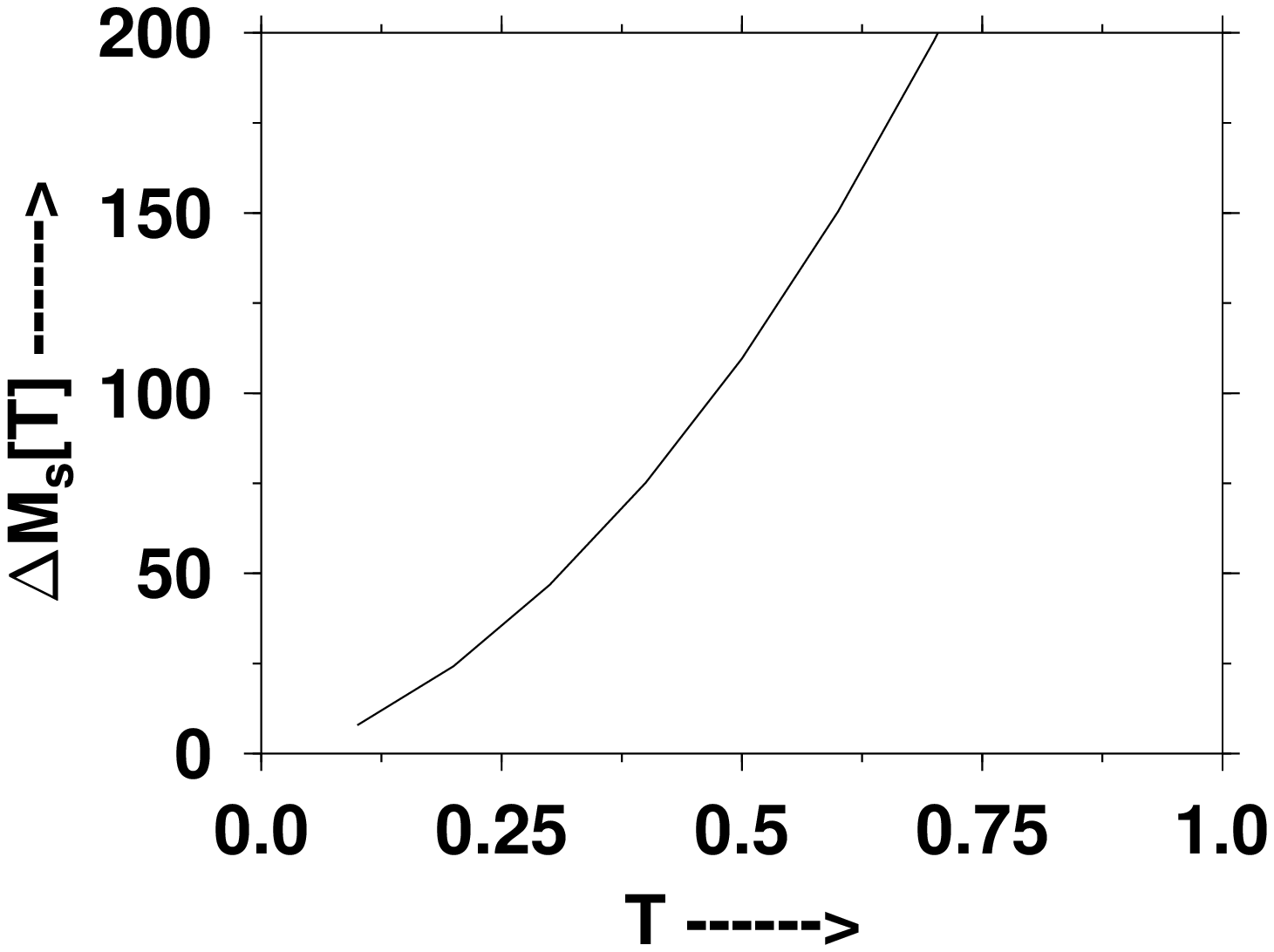}\hglue 0.35 truecm}

\vglue 0.8 truecm
\noindent {\bf Fig.\ 2:}
{{{Here for dimensionalities $d=2$ (left panel) and $d=3$ (right panel),
respectively, the  integral $\Delta M_s(T):=M_s(0)-M_s(T)\propto$
\beqa
&&\int\limits_{k=0}^{k=\pi}
\frac{{\rm d}^dk}{
\exp \{T^{-1}\sqrt{
[2\cdot (1-\cos k )+C_b]\cdot[2\cdot (1-\cos k)
+k\cdot\tilde C_u\cdot\frac{2k_y^2}{k^2}]}
\}
-1 
}
\nonumber\cr  &&\times \frac{2\cdot (1-\cos k) 
+\frac{C_B}{2}+k\cdot\tilde C_u\cdot\frac{k_y^2}{k^2}}
 {\sqrt{[2\cdot (1-\cos  k )
+C_b]\cdot[2\cdot (1-\cos k ) +k\cdot\tilde C_u\cdot\frac{2k_y^2}{k^2}] }}\nonumber
\eeqa is presented
over the reduced Kelvin temperature $T$.

This plot represents schematically (in the left panel) the temperature
dependence of the magnetisation of an in-plane magnetised ultrathin film
under the influence of a typical exchange interaction $\eps =2\cdot (1-\cos
k)$, which behaves for small $k$ as $\eps =k^2$, while the reduced dipolar
anisotropy $\propto 4\pi M_s$ is represented by $C_b=0.01$, and, finally, a
$k$-dependent effective in-plane uniaxial anisotropy corresponding to the
last term in Eqn.\ (\ref{eqEffectiveAnisotropy}), $\eps(k)\propto\sqrt
{...+(4 \pi M_s)^2\cdot (k\cdot t_h/2)\cdot (k_y^2/k^2)}$, is represented by
the terms $\propto k\cdot\tilde C_u$, where $\tilde C_u=0.001$ is chosen.
The results of the left panel correspond roughly to the 'D\"oring behaviour' $
70\cdot T\cdot
\ln\frac {T}{0.05}$, which necessitates for $d=2$ an additional interaction
factor $F_k(T)\propto T^{1/2}$ to obtain a $T^{3/2}$-behaviour, whereas the
'$d=3$- results' of the right panel (from which the difference can be seen)
would yield just this behaviour.

}}}

\newpage

\hglue 2.5 truecm
\epsfxsize=7cm
\epsfbox{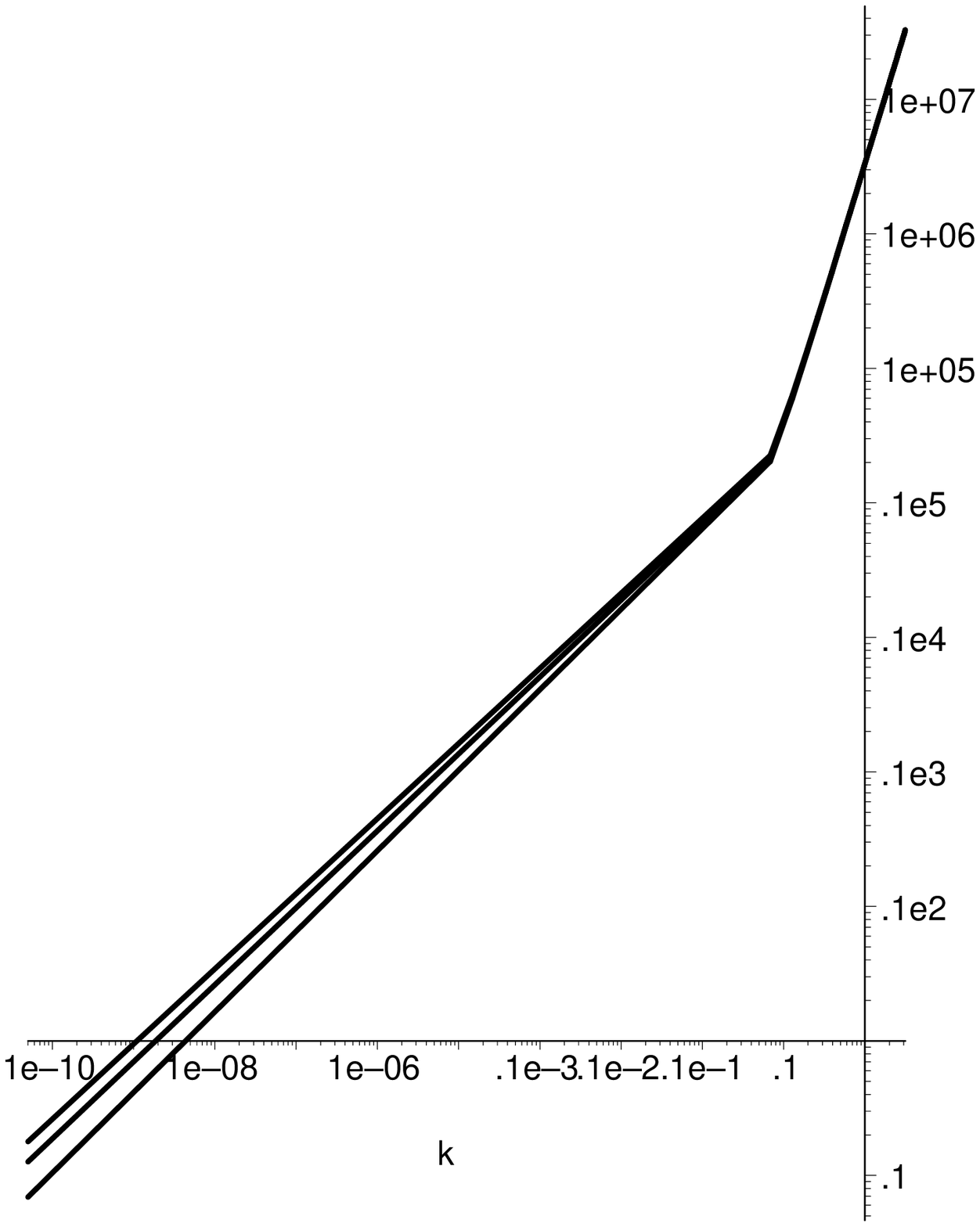}
\vglue 0.3 truecm

\noindent {\bf Fig.\ 3:}
{{{In a double-logarithmic plot the energy $\eps(\vec k)$ (up to a constant
factor) is presented over $k$ for $\vec k =(0,k,0)$, for material parameters
of permalloy as in the preceding figure, where $k$ is measured in
$(nm)^{-1}$. The thicknesses of the ultrathin films considered, are 4 $nm$,
2 $nm$, and $0.6\,\, nm$, respectively, corresponding to the highest, medium
rsp.\ the lowest
 curve. 

Note  the  sharp crossover around $k=k_c\equiv\pi t_h/l_e^2\cong 0.1\,\,
nm^{-1}$.  For shorter wavelengths (larger $k$) the different thicknesses
can no longer be distinguished, since the exchange interaction dominates.
 }}}

\newpage


\hglue 3.5 truecm
\epsfxsize=5cm
\epsfbox{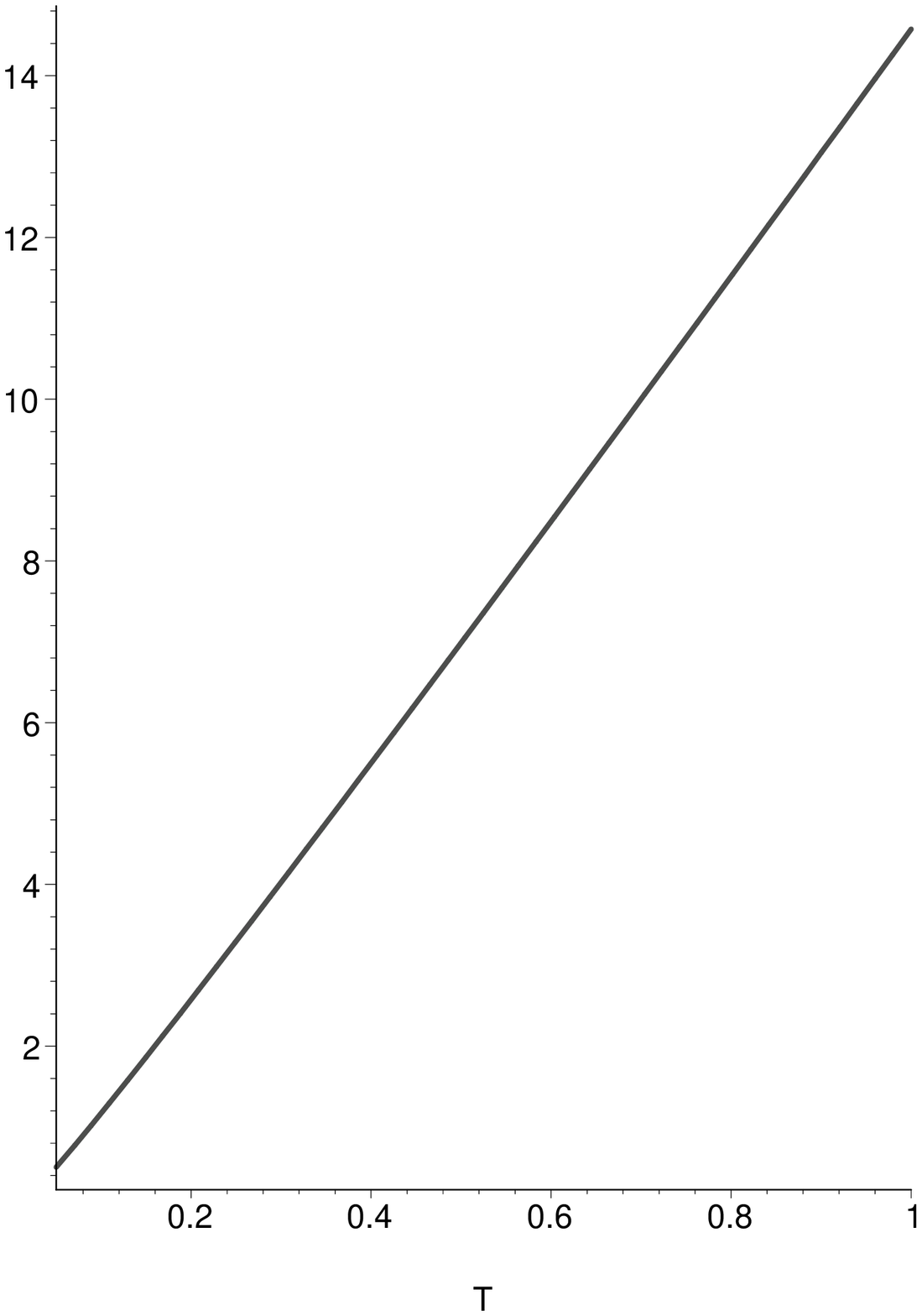}
\vglue 0.3 truecm

\noindent {\bf Fig.\ 4:}
{{{ For $C_b=0.01$, as in Fig.2, the integral
$M_s(0)-M_s(T)\propto$
\beqa
&&\int\limits_{k=0}^{k=\pi} 
\frac{\,\,{\rm d}k}{\exp \{T^{-1}[\,\,2\cdot (1-\cos
 k )+C_b]-1 \}}\nonumber\eeqa is presented
over the reduced Kelvin temperature $T$. The results should be valid for
$d=1$.

}}}

\begin{thebibliography}{99}
\bibitem{REM4} The Curie temperature of a Heisenberg
model with a large value of $s$ is $\propto s^2$, whereas the largest magnon
energy is only $\propto s$. Here we only need $T_1\ll T_2$, where
$T_2$ (rsp.\ $T_1$) is the upper (rsp.\ lower) crossover temperature
of the 'Bloch region' $T_1 < T < T_2$.
\end{thebibliography}
\end{document}